\documentclass[prb,preprint,showpacs,preprintnumbers,amsmath,amssymb]{revtex4}

\usepackage{graphicx,color,psfrag}
\usepackage{dcolumn}
\usepackage{bm}

\begin{document}

\title{Strong extinction of a far-field laser beam by a single quantum dot}
\author{A. N. Vamivakas$^1$}

\author{M. Atat\"{u}re$^{2,3}$}
\author{J. Dreiser$^2$}%
\author{S. T. Yilmaz$^2$}%
\author{A. Badolato$^2$}%
\author{A. K. Swan$^1$}
\author{B. B. Goldberg$^{1,4}$}
\author{A. Imamo\u{g}lu$^{2}$}
\email{imamoglu@phys.ethz.ch}
\author{M. S. \"{U}nl\"{u}$^{1,4}$}
\email{selim@bu.edu}

\affiliation{%
$^1$Department of Electrical and Computer Engineering, Boston University, Boston, MA 02215, USA.\\
$^2$Institute of Quantum Electronics, ETH Z{\" u}rich, HPT G10, 8093 Z{\" u}rich, Switzerland.\\
$^3$Cavendish Laboratory, University of Cambridge, JJ Thomson Avenue, Cambridge CB3 0HE, UK.\\
$^4$Department of Physics, Boston University, Boston, MA 02215, USA}

\date{\today}
\begin{abstract}
    Through the utilization of index-matched GaAs immersion lens
    techniques we demonstrate a record extinction (12\%) of a
    far-field focused laser by a single InAs/GaAs quantum dot.  This contrast level enables us to report for the first time resonant laser transmission spectroscopy on a single InAs/GaAs quantum dot without the need for phase-sensitive lock-in detection.
\end{abstract}

%
\maketitle

The interaction of light and matter at the nanometer length scale is
at the heart of the burgeoning field of nanophotonics \cite{NanoOpt}.
A prerequisite to the efficient optical manipulation and interrogation
of a nanophotonic system is the ability to couple external
electromagnetic radiation to and from the system of interest.  The
problem of light coupling is exacerbated if the physical system of
interest is buried beneath high refractive index-air boundaries
\cite{Gold} as is the case for most solid-state systems.  Here we
demonstrate, through the utilization of index-matched 
GaAs immersion lens techniques \cite{Kino,Ipp1} a record extinction (12\%) of a far-field focused laser by a single InAs/GaAs quantum dot.  The lens system provides a seven-fold improvement in extinction over the planar interface. This contrast level enables us to report for the first time resonant laser transmission spectroscopy on a single InAs/GaAs quantum dot without the need for phase-sensitive lock-in detection.  

The extinction cross-section quantifies the ability of a quantum
mechanical 2-level system to extinguish the energy of a monochromatic
plane wave.  At resonance between the incoming plane wave and the
2-level system transition, the extinction cross-section is
$\sigma_{res}=3/2 \pi \lambda^{2}$ \cite{OptCoh}, provided that the line
  broadening is solely due to spontaneous emission on the probed
  transition.  A comparison of the far-field diffraction limited
  focused beam area $A_f = \pi (\lambda/4NA)^2$ with the on-resonance cross-section $\sigma_{res}$
suggests that a dipole-like 2-level system should be able to strongly
extinguish the illuminating beam energy with a modest focusing
objective numerical aperture of $NA = 0.65$.  A more careful
analysis shows that strong extinction would require a good matching of
the focused Gaussian laser beam with the dipole emission profile of
the two-level emitter \cite{Enk1,Enk2}.

Experimental studies reveal 6\% extinction \cite{note1}  of light from the 100 nm aperture of a near-field tip onto a single dibenzanthanthrene molecule \cite{Sand1}. The aluminum coated tapered fiber localized the light field beyond the diffraction limit and enhanced the near-field coupling to the molecule, but the presence of the tip influences the isolated molecule's optical response.  Another physical realization of the 2-level quantum mechanical system is the ground state exciton X$^0$ in InAs quantum dots (QDs) buried in a GaAs host \cite{Hoge1}.   In this recent work the optical coupling to the QD was limited by the dielectric boundary formed between the GaAs host matrix and the surrounding medium. Conventional far-field measurements through planar boundaries limit the focused beam spot diameter to about half the vacuum wavelength resulting in measurements with a maximum extinction by the quantum dot of 1.7\% and require lock-in detection to remove both laser and electrical noise. 
	
Solid immersion technology enhances coupling of light to and from molecular beam epitaxy grown QDs buried within the planar structure of a high index GaAs host matrix. Refraction at the planar vacuum-GaAs dielectric boundary limits the collection/focusing angle for light to a maximum of ~17 degrees (nGaAs = 3.475 at 960 nm.). An index-matched numerical aperture increasing micro-lens (NAIL) \cite{Ipp1} (also referred to as a SIL, depending on configuration \cite{Ipp2}) alters the planar boundary geometry providing maximum coupling. Glass and index-matched GaAs SIL/NAIL techniques have been employed in micro-photoluminescence studies of single InAs/GaAs QDs \cite{Liu,Zwil1} and recently an epitaxy-side glass SIL has been used in a resonant scattering measurement \cite{Gera1}, demonstrating reduction of the laser spot area and stronger interaction with single QDs. In this work we introduce both an epitaxy-side GaAs SIL and a substrate-side GaAs NAIL into the resonant light scattering measurement system so that the QD is accessible optically from nearly the full 4$\pi$ of solid angle. The top SIL reduces the focal spot area, while the bottom NAIL improves light collection. We demonstrate record far-field extinction of a focused laser beam, and typical QD extinction is visible even using a dc power-meter without the need for phase-sensitive lock-in detection to monitor the light resonantly scattered from a single QD.  

The experimental setup is illustrated in Fig. 1(a), with the SIL/QD/NAIL sample assembly sandwich illustrated in Fig. 1(b). Two ceramic pieces auto-align the SIL and NAIL on the optical axis of the sample assembly, clamp the lens sample assembly into position providing optical contact between the QD sample and both lenses. To compare the extinction enhancement, measurements have been made with two types of assemblies.  Sample assembly 1 (SA1, Fig. 1(b)) consists of a 1.61-mm radius GaAs SIL in optical contact with the epitaxial QD sample surface and a 1.61-mm radius GaAs SIL (also referred to as a NAIL) in optical contact with the substrate side of the QD sample.  A second assembly (SA2) consisted of a single 1-mm radius GaAs SIL in optical contact with the QD sample epitaxial surface and no lens mounted on the substrate side of the QD sample.     

The detected field in transmission measurements is a coherent superposition of the field scattered by the X$^0$ exciton and the transmitted laser light that drives the transition. The differential transmission is related to both the QD extinction cross-section and the numerical aperture of the light focusing optics as \cite{Karr1}:

\begin{equation}
\label{DT}
\frac{\Delta T}{T}= 1 - \frac{\sigma(\Omega,\omega_L)}{A_f}G
\end{equation}

\noindent where we assume the laser is a monochromatic plane wave,
$\sigma$ is the QD extinction cross-section, $\Omega$ is the Rabi
frequency, $\omega_L$ is the driving laser frequency, $G<1$ is a
factor taking into account the fact that the focused Gaussian and the
dipole emission profiles have an imperfect overlap \cite{Enk2} and $A_f$ is the
focused beam area approximately equal to $\pi(\lambda/4NA)^2$.  Eq. (1) is valid
provided that $\Delta T \thicksim T$, which is satisfied in all such
experiments to-date. In Fig. 2 we present resonant scattering data
from the X$^0$ transition in single InAs/GaAs QDs measured with a
lock-in detection technique by amplitude modulation of the gate
voltage \cite{Karr2}.   In Fig. 2(a) we compare the power saturation curves
for three different QDs with two different external numerical aperture
objectives.  The red diamonds fit to a contrast of 1.70 $\pm$ 0.15 \%
for the best X$^0$ data measured in the standard planar interface
sample configuration with a 0.65 NA external objective. The blue
triangles in Fig. 2(a) fit to a saturation curve with a contrast of
6.77 $\pm$ 0.12 \% for an average QD in the SA2 configuration, and the
orange circles fit to a contrast of 8.66 $\pm$ 0.24 \% for the best QD
in the SA1 configuration.  For the 12 QDs measured in both configurations SA1 and SA2 we found an average contrast of 6.65$\pm$2.4\%.  The best case improvement of $\backsim$7 measured in the SA2 configuration, as determined from comparing to the planar contrast [red diamonds in Fig 2(a)], is plotted in Fig. 2(a) with purple circles. Remarkably, this particular QD extinguishes $\backsim$12\% of the illuminating beam energy. Figure 2(b) displays one of the three laser scans performed to obtain the highlighted data point in Fig. 2(a) at a fixed optical power of 0.02 nW.

To demonstrate our improved ability to couple light to the InAs QD we
eliminated the phase-sensitive lock-in detection electronics and
measured the resonantly scattered light directly as we tuned the X$^0$
resonance with applied gate voltage.  Figure 3 presents transmission
as a function of laser power demonstrating, the 2-level atomic system
signature of power broadening \cite{Stroud} for the X$^0$ transition for the
dot with the average contrast (triangle data) in Fig 2(a).  This data
is the first far-field resonant light scattering experiment reported
on a solid-state-based 2-level system without the need for phase
sensitive lock-in detection.  Further, we measure at least 7\%
extinction of the illuminating laser on the dc power-meter when the
laser power is 0.03 nW (saturation power is 0.206 nW). At these optical powers the QD sees a photon about once every 10 radiative lifetimes and we detect an easily observable extinction contrast in the far-field measured light, despite no active noise reduction in the measurement technique.

The maximum number of photons a 2-level system can scatter is fixed
due to saturation. Although we do increase the measured contrast by a
factor of 7 in the SIL/QD configuration (SA2) we want to stress that
we do not change the extinction cross-section of the dot, but only the
input power necessary to scatter a fixed number of photons.   The
effect of the epitaxy-side SIL is to decrease the focal spot area of
the illuminating laser and therefore reduce the necessary input power
to drive the dot to saturation; it does not change the optical
response of the QD in the linear (below-saturation) regime.  However,
the substrate-side planar GaAs interface causes total internal
reflection reducing the laser and dipole field intensity at the
detection plane. The SIL/QD/NAIL configuration (SA1) is the more
suitable system for future experiments as this removes the effects of
total internal refraction of the laser and dipole field and results in
an increased signal at the detection plane. 

The theoretical enhancement in contrast for small NA is approximately
the square of the GaAs refractive index which at 960 nm is
$\backsim$12.1 for two index matched materials. We measure more modest
values of contrast enhancement, most likely limited by the quality of
the optical contact between the epitaxy-side SIL and the QD sample.
In our geometry, the metallic top Schottky contact inhibits the
optical contact between the SIL and QD sample and the vacuum-GaAs
boundaries form a Fabry-Perot cavity that, depending on the QD
transition wavelength, can enhance or suppress the QD emission
influencing the measured contrast \cite{Karr1}.  
 
Immersion lens techniques have markedly improved our ability to couple
light to and from single buried solid state-based light emitters
yielding a record 12\% extinction of strongly focused light. The next
step in this particular direction will be to improve the spatial mode
matching between the laser and dipole fields [G in Eq. (1)]. Further,
the increase in far-field accessible object space solid angle suggests
tailoring the excitation beam amplitude distribution to yield specific
vectorial field distributions in the object space, opening the
possibility of quantum nanophotonics with engineered vector fields.  A
compelling interest in the fundamental limit of strongly focused light
extinction by a 2-level system is the efficient coupling of light to
qubits realized in the physical degrees of freedom of either single
atoms or semiconductor QDs \cite{Enk1}.  Our improved ability to couple light
to the QD has removed the need for phase-sensitive electronic
detection in a standard resonant scattering experiment.  Directly
monitoring light scattered from an InAs QD without a lock-in opens the
door for real-time measurement and control of single QDs without the
need for voltage or laser modulation. 
Additionally, the recent demonstration of time-averaged single spin
measurement by means of Faraday rotation estimates that incorporation
of SIL/NAIL into a resonant scattering experiment can make accessible
an experimental regime in which the dynamics of a single electron spin
can be monitored without significant back-action on the spin degree of
freedom induced by the measuring optical field \cite{Ata1,Aws1}  -- a technical necessity for the implementation of optical quantum information processing protocols \cite{Imm1,DiV1}.


\begin{acknowledgments}
The work presented here was supported by NCCR Quantum Photonics, the Air Force Office of Scientific Research under grant MURI F-49620-1-0379 and by the National Science Foundation under grant NIRT ECS-0210752. The authors are grateful to K. Karrai for bringing up the interface related cavity effect and J. Dupuis for assistance in modeling the detection optical system.
\end{acknowledgments}

\newpage

\begin{figure}[t]
\begin{center}
\includegraphics[width=5in]{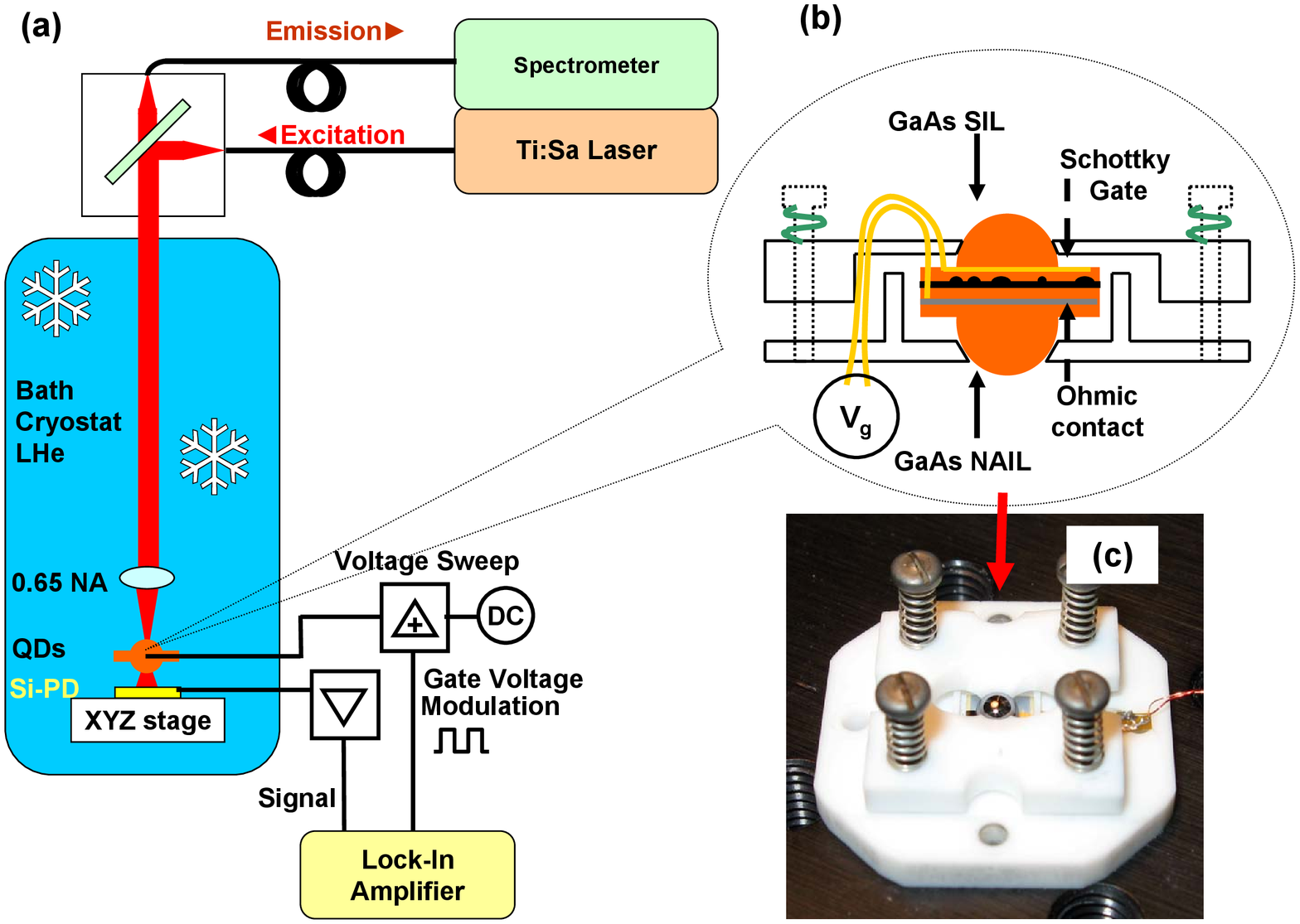}
\end{center}
\caption{a) An illustration of the experimental apparatus used for
  both micro-photoluminescence and resonant scattering measurements. 
b) A schematic of the sample-NAIL/SIL assembly.  c) A picture of the contacted sample assembly illustrated in b.
}
\label{Fig1}
\end{figure}

\begin{figure}[t]
\begin{center}
\includegraphics[width=5in]{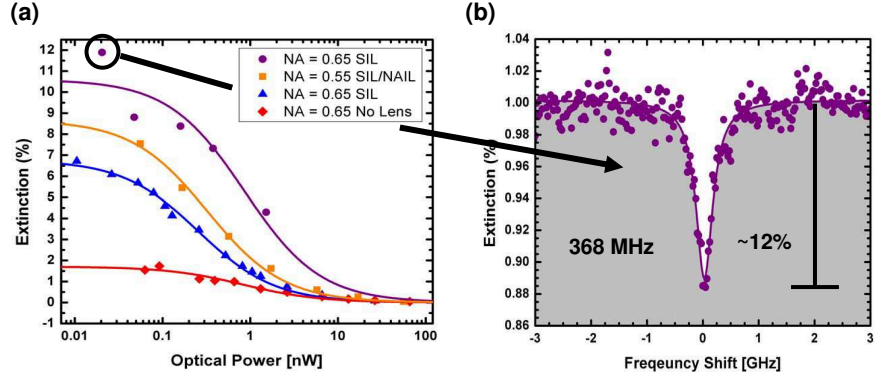}
\end{center}
\caption{a) We fixed the diode laser frequency to be commensurate with the X$^0$ transition frequency and measured the strength of the scattered light signal as a function of incident laser power.  To evaluate the true extinction ratio of the QD transition, we take the raw measured lock-in signal power, divide by the incident laser power, and rescale the measured saturation curves by an experimentally obtained correction factor (2.25 or 2 depending on the lock-in amplifier) to account for the reduction in contrast resulting from the lock-in measurement technique.  Each data point for the four saturation curves is the average of three separate measurements.  The red diamonds are the saturation data with no immersion lens incorporated into the setup.  The blue triangles are data measured on the average quantum dot with a r = 1 mm GaAs SIL and a 0.65 numerical aperture objective.  The purple circle data is the best dot we encountered with the r = 1 mm GaAs SIL.  The orange square data is taken with both the r = 1.62 mm GaAs SIL and NAIL with a 0.55 numerical aperture objective.  All data points are averages from 3 measurements each recorded with a lock-in time constant of 100 ms.  b,  The best linescan recorded for the lowest power point on the purple power saturation curve in a.  The measured contrast is 12\% and the linewidth is 368 MHz (1.47 $\mu$eV).       }
\label{Fig2}
\end{figure}

\begin{figure}[t]
\begin{center}
\includegraphics[width=5in]{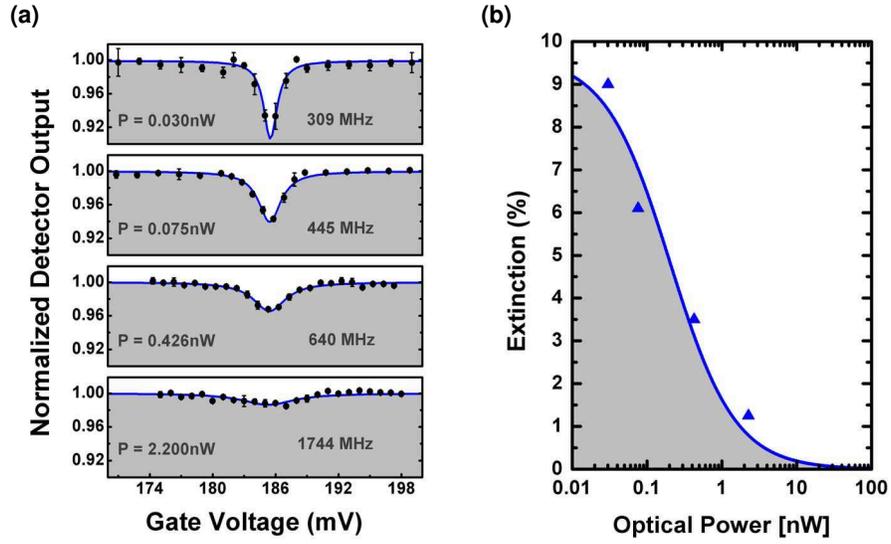}
\end{center}
\caption{a) Linescans as a function of incident laser power to demonstrate power broadening of the QD X$^0$ transition.  The linescans are recorded for laser powers of 0.030, 0.075, 0.426, and 2.200 nW (ordered from top linescan to bottom linescan).  The lowest power linescan fits to a linewidth of 309 MHz (1.27 $\mu$eV).  The lower left inset is the power at which the linescan is recorded and the lower right inset is the measured FWHM.  b)  The saturation curve for QD studied in a.  }
\label{Fig3}
\end{figure}


\begin{thebibliography}{21}
\bibitem[{not()}]{note1}
\bibinfo{note}{Upon completion of this work we became aware that 10\%
  extinction has been measured on this same molecule in the far-field.}
\end{thebibliography}
\end{document}